\documentclass[a4paper,10pt,twocolumn]{article}
\usepackage{amsmath,amsfonts}
\usepackage{authblk}
\usepackage{algorithmic}
\usepackage{algorithm}
\usepackage{array}
\usepackage{caption}
\usepackage{subfig}
\usepackage{textcomp}
\usepackage{stfloats}
\usepackage{url}
\usepackage{verbatim}
\usepackage{hyperref}
\usepackage{graphicx}
\usepackage{cite}
\usepackage[table, dvipsnames]{xcolor}
\usepackage{booktabs}
\usepackage{multirow}
\usepackage{listings}
\usepackage{float}
\usepackage{authblk}
\usepackage{xcolor}
\usepackage{tikz}
\usepackage{eso-pic}
\usepackage[utf8]{inputenc}
\usepackage{amsmath, amssymb}
\usepackage{graphicx}
\usepackage{tikz}
\usepackage{pgfplots}
\usepackage{algorithm}
\usepackage{algorithmic}
\usepackage{enumitem}
\usepackage{multirow}
\usepackage{tabularx}
\usepackage{url}
\usepackage{cite}
\usepackage{hyperref}

\pgfplotsset{compat=1.15}
\title{\LARGE \bf Advancing Face-to-Face Emotion Communication: A Multimodal Dataset (AFFEC)}

\author{%
  Meisam J. Sekiavandi$^{*,1,3}$,
  Laurits Dixen$^{*,1,3}$,
  Jostein Fimland$^{1,3}$,
  Sree Keerthi Desu$^{2}$,
  Antonia-Bianca Zserai$^{1}$,
  Ye Sul Lee$^{1}$,
  Maria Barrett$^{1}$,
  Paolo Burreli$^{1,3}$\\
  
  $^{1}$IT University of Copenhagen, 
  $^{2}$Technical University of Denmark\\
  $^{3}$Pioneer Centre for Artificial Intelligence, $^{*}$Equal contributions, 
}
\AddToShipoutPictureBG*{%
  \begin{tikzpicture}[remember picture,overlay]
    \node[
      rotate=45,
      anchor=north,        % align the top of the text with the page top
      yshift=-4cm,  
      xshift=7.5cm,% move it 1 cm down from the very top edge
      text=white,         % white text
      fill=gray,         % black background
      inner sep=3pt,      % padding around the text
      text=gray!60,
      scale=1.5
    ]
    at (current page.south) {\textbf{ \quad \quad not peer-reviewed \quad \quad 
 }};
  \end{tikzpicture}%
}

\date{} 
\begin{document}

% Change footnote numbering to symbols
\renewcommand{\thefootnote}{\fnsymbol{footnote}}

\maketitle

\begin{abstract}
Emotion recognition has the potential to play a pivotal role in enhancing human-computer interaction by enabling systems to accurately interpret and respond to human affect. Yet, capturing emotions in face-to-face contexts remains challenging due to subtle nonverbal cues, variations in personal traits, and the real-time dynamics of genuine interactions. Existing emotion recognition datasets often rely on limited modalities or controlled conditions, thereby missing the richness and variability found in real-world scenarios.

In this work, we introduce \emph{Advancing Face-to-Face Emotion Communication} (AFFEC), a multimodal dataset designed to address these gaps. AFFEC encompasses 84 simulated emotional dialogues across six distinct emotions, recorded from 73 participants over 5,000+ trials and annotated with more than 20,000 labels. It integrates electroencephalography (EEG), eye-tracking, galvanic skin response (GSR), facial videos, and Big Five personality assessments. Crucially, AFFEC explicitly distinguishes between \emph{felt} emotions (the participant’s internal affect) and \emph{perceived} emotions (the observer’s interpretation of the stimulus). 

Baseline analyses—spanning unimodal features and straightforward multimodal fusion—demonstrate that even minimal processing yields classification performance significantly above chance, especially for arousal. Incorporating personality traits further improves predictions of felt emotions, highlighting the importance of individual differences. By bridging controlled experimentation with more realistic face-to-face stimuli, AFFEC offers a unique resource for researchers aiming to develop context-sensitive, adaptive, and personalized emotion recognition models.
\end{abstract}

% \begin{IEEEkeywords}
% Emotion recognition, multimodal dataset, human-agent interaction, EEG, eye-tracking, affective computing.
% \end{IEEEkeywords}

\section{Introduction}
Human communication is inherently emotional, relying on both verbal and non-verbal cues to convey meaning and build connections. With the rapid rise of Human-Agent Interaction (HAI) and Social Robotics (SR), there is an increasing demand for agents capable of understanding and responding to human emotions. Such systems are critical for enhancing interactions in healthcare, education, customer service, and entertainment \cite{mohammed2020survey}. The ultimate goal is to develop socially adept agents that consider both the cognitive and emotional states of their human counterparts, ensuring safe, engaging, and adaptive interactions.

Despite significant advancements, current emotion recognition systems struggle to capture the full complexity of real-world human-agent interactions. Many approaches rely heavily on verbal cues while neglecting subtle non-verbal signals—such as facial expressions, eye movements, and physiological responses—that are essential for accurately interpreting emotions \cite{cakir_reviewing_2023}. Moreover, existing datasets are often constrained by static or artificial stimuli, limited modality diversity, and a lack of consideration for idiosyncratic dispositions (e.g. as expressed by personality traits) that critically influence emotional expression and perception \cite{mohammadi2020multi, hosseini2023personality, khare2024emotion, 9395500, park2020k}.

\begin{figure}[H]
\centering
\includegraphics[width=0.3\textwidth]{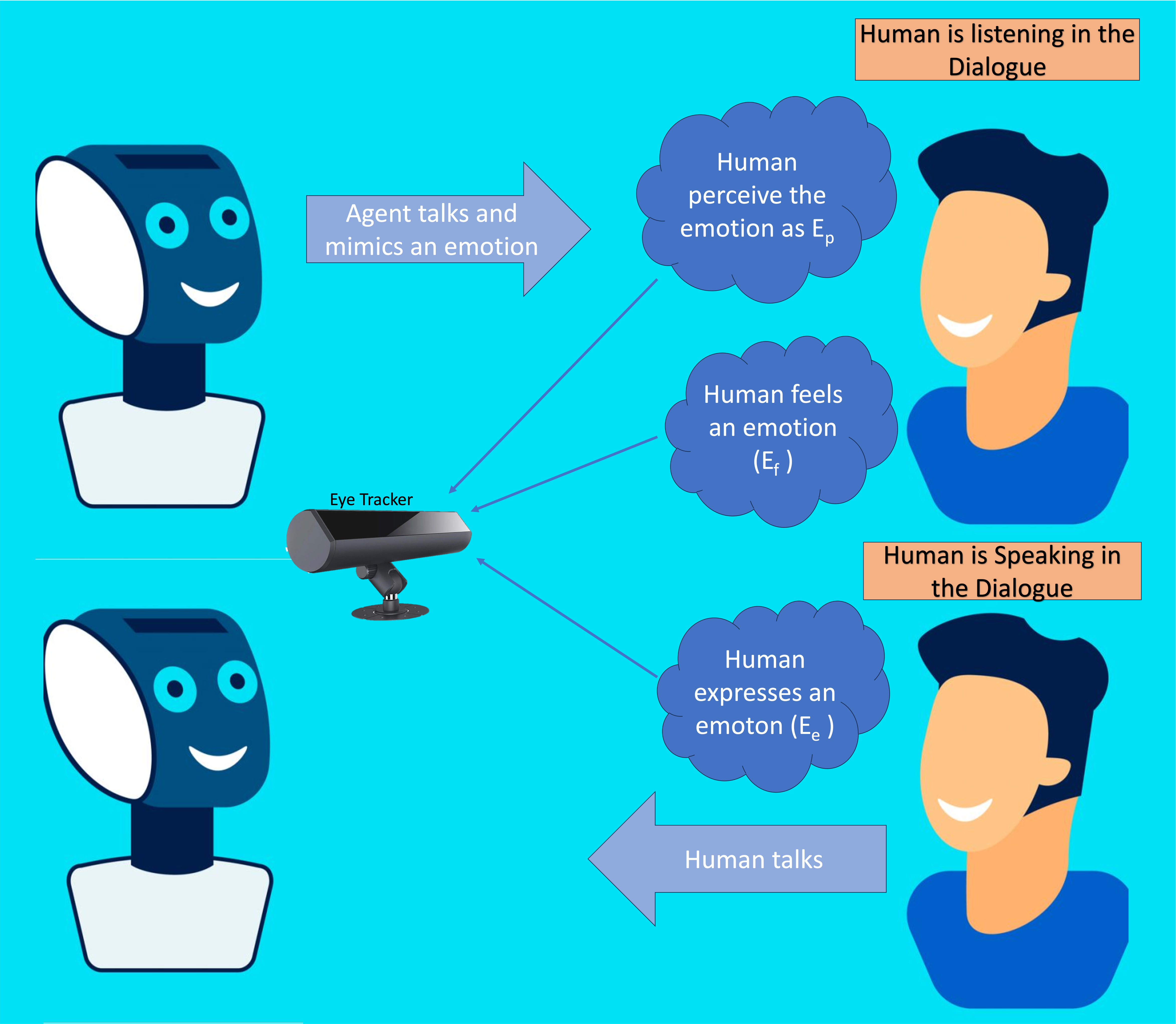}
\caption{An illustrative overview of the three types of emotions examined in this study: Expressed emotion ($E_e$), Perceived emotion ($E_p$), and Felt emotion ($E_f$).}
\label{fig:HAI}
\end{figure}

The importance of emotions in HAI is underscored by studies demonstrating that users respond more positively to virtual agents that recognise and adapt to their emotional states \cite{yang_emotion_2021, cao_analysis_2021}. To achieve this, it is necessary to differentiate among \textit{Expressed Emotions} ($E_e$), \textit{Perceived Emotions} ($E_p$), and \textit{Felt Emotions} ($E_f$). When encountering emotional stimuli, individuals not only perceive the emotion ($E_p$) but also experience an internal affective response ($E_f$). Figure~\ref{fig:HAI} illustrates the interplay among these categories \cite{horvath2022felt, saarimaki_cerebral_2023}.

To address these challenges, we introduce the \textit{Advancing Face-to-Face Emotion Communication} (AFFEC) dataset. AFFEC is designed to capture the dynamic complexity of face-to-face emotional interactions by integrating multiple modalities—EEG, eye-tracking, GSR, facial movements, and personality assessments—with explicit labels for both perceived and felt emotions. This rich dataset not only facilitates a more comprehensive understanding of emotional processes but also enables the exploration of how individual differences modulate affective experiences.

\subsection{Contributions of this Work}
The main contributions of this work are as follows:
\begin{itemize}
    \item \textbf{Introduction of the AFFEC Dataset:} A multimodal dataset capturing 84 simulated dialogues in six distinct emotions from 73 participants, comprising over 5,000 trials and 20,000+ labels, including EEG, eye tracking, GSR, facial, and personality data.
    \item \textbf{Explicit Differentiation of Emotional Categories:} AFFEC uniquely distinguishes between expressed ($E_e$), perceived ($E_p$), and felt emotions ($E_f$), enabling nuanced analysis of human-agent emotional exchanges.
    \item \textbf{Baseline Analyses:} We present baseline models using unimodal and multimodal features, demonstrating that even minimal processing yields performance levels well above chance, particularly for arousal prediction.
    \item \textbf{Incorporation of Individual Differences:} By integrating personality assessments, AFFEC supports the exploration of how stable individual traits influence emotional perception and expression, paving the way for personalized affective computing.
    \item \textbf{Advancing HAI and SR:} AFFEC provides a robust foundation for developing virtual agents and robots that recognize and respond to the dynamic, context-dependent nature of human emotions.
\end{itemize}

Overall, AFFEC addresses critical gaps in emotion recognition research and offers a valuable resource for advancing the fields of affective computing, human-agent interaction, and social robotics.

\section{Related Work}
Emotion recognition is a central topic in affective computing, with the goal of enabling machines to detect, interpret, and adapt to human emotions. Over the past decades, extensive research has explored various modalities—such as facial expressions, speech, and physiological signals—and a range of models to improve emotion detection accuracy and robustness. However, many existing studies rely on simplified emotion models and controlled settings, limiting their ecological validity and personalization.

\subsection{Emotion Models}
Emotion recognition research typically employs discrete or dimensional models. Discrete models categorise emotions into fundamental classes—often using the Big Six (anger, disgust, fear, joy, sadness, surprise)~\cite{ekman2005argument}—while dimensional models represent emotions continuously along axes like arousal and valence (and sometimes dominance)~\cite{mehrabian1996pleasure}. Although discrete models offer intuitive classifications, they may oversimplify the richness of spontaneous expressions in the real world~\cite{siegert2011appropriate}.

\subsection{Multimodal Emotion Recognition}
Recent work has increasingly focused on multimodal emotion recognition to address the limitations of unimodal approaches. For example, Kollias et al.~\cite{kollias2022} employed a multi-task learning framework to jointly predict valence, arousal, facial expressions, and action units from audio-visual inputs, demonstrating significant performance gains. Reviews by Li et al.~\cite{li2024} and Zhang et al.~\cite{zhang2023} further underscore that combining facial, speech, and physiological signals outperforms single-modality systems. However, many studies focus primarily on arousal and valence metrics, potentially neglecting the full complexity of human emotion.

\subsection{Stimuli, Participant, and Temporal Considerations}
Traditional emotion recognition research often relies on emotionally charged video clips, music, or data from actors~\cite{park2024, chen2024}. Although these controlled elicitation methods yield reproducible ground truths, they may fail to capture the subtlety and spontaneity inherent in everyday interactions. Moreover, many studies use static or short-duration stimuli that overlook the temporal evolution of emotions\cite{seikavandi2023gaze,j2024modeling}. As noted by Wang et al.~\cite{wang2023}, integrating temporal dynamics is essential for accurately modeling how emotions unfold over time.

\subsection{Personality and Emotion Recognition}
Individual differences, particularly personality traits, play a significant role in shaping emotional perception and expression. For instance, neuroticism is associated with heightened negative affect and arousal~\cite{kehoe2012personality}, whereas extraversion correlates with increased positive emotions~\cite{zautra2005dynamic} and openness enhances receptivity to diverse emotional stimuli~\cite{costa1980influence}. Incorporating personality measures into emotion recognition systems can lead to more personalised and adaptive models.

\subsection{Notable Emotion Recognition Datasets}
Several multimodal datasets have advanced the field:
\begin{itemize}
    \item \textbf{DECAF} \cite{abadi2015decaf} integrates MEG, hEOG, ECG, tEMG, video, and audio from music videos and film clips, capturing valence, arousal, and dominance.
    \item \textbf{SEED-VII} \cite{jiang2024seed} uses EEG and eye tracking to record responses to video clips designed to evoke six basic emotions and a neutral state.
    \item \textbf{DREAMER} \cite{katsigiannis2018dreamer} employs low-cost EEG and ECG sensors to capture affective responses to film clips.
    \item \textbf{EAV} \cite{lee2024eav} combines EEG, audio, and video recordings in scripted conversational settings.
    \item \textbf{AMIGOS} \cite{miranda2021amigos} collects EEG, ECG, GSR, and audiovisual data during individual and group emotional experiences.
    \item \textbf{AffectNet} \cite{mollahosseini2019affectnet} provides over 400,000 annotated images for facial expression analysis.
    \item \textbf{K-EmoCon} \cite{park2020kemocon} captures continuous emotional states during naturalistic conversations using EEG, ECG, physiological signals, and video.
    \item \textbf{WESAD} \cite{schmidt2018wesad} focuses on wearable stress and affect detection via multiple physiological signals.
    \item \textbf{MAHNOB-HCI} \cite{soleymani2012multimodal} integrates facial videos, audio, eye gaze, and various physiological signals.
    \item \textbf{ASCERTAIN} \cite{subramanian2018ascertain} combines EEG, ECG, and GSR with self-reports and personality assessments.
    \item \textbf{MGEED} \cite{wang2024mgeed} provides rich annotations from EEG, ECG, facial optomyography (OMG), video and depth maps.
    \item \textbf{DEAP} \cite{koelstra2012deap} integrates EEG, GSR, and facial videos from music videos.
    \item  \textbf{SEMAINE} \cite{mollahosseini2019affectnet} focusses on textual and dyadic conversational scenarios.
\end{itemize}
Table~\ref{tab:dataset_comparison} summarises these datasets alongside AFFEC, highlighting AFFEC’s unique integration of multimodal signals and its explicit differentiation between felt and perceived emotions.

\subsection{Limitations and the AFFEC Contribution}
While existing datasets have advanced the field, they often suffer from:
\begin{itemize}
    \item \textbf{Limited Multimodal Diversity:} Many exclude subtle non-verbal cues such as gaze and personality assessments.
    \item \textbf{Static or Contrived Stimuli:} Reliance on staged expressions limits ecological validity.
    \item \textbf{Lack of Emotion Differentiation:} Few datasets distinguish between felt and perceived emotions.
    \item \textbf{Participant Homogeneity:} Limited demographic diversity undermines generalisability.
    \item \textbf{Neglected Temporal Dynamics:} Most datasets fail to capture the evolution of emotions over time.
\end{itemize}

AFFEC directly addresses these gaps by integrating EEG, eye tracking, GSR, facial movements, and personality assessments. It uniquely differentiates between expressed, perceived, and felt emotions using naturalistic stimuli from non-actor participants and continuous recordings, thereby providing a more comprehensive foundation for developing adaptive, context-aware emotion recognition models.

\begin{table*}[H]
\renewcommand{\arraystretch}{1.5}
%\small
\centering
\caption{Comparison of SOTA Multimodal Emotion Recognition Datasets and AFFEC}
\label{tab:dataset_comparison}
\begin{tabular}{p{3cm}p{4cm}p{2cm}p{4cm}p{3cm}}
\hline
\textbf{Dataset} & \textbf{Modalities} & \textbf{Participants} & \textbf{Stimuli} & \textbf{Emotions} \\
\hline
DECAF \cite{abadi2015decaf} & MEG, hEOG, ECG, tEMG, Video, Audio & 30 & Music videos, film clips & Valence, Arousal, Dominance \\
SEED-VII \cite{jiang2024seed} & EEG, Eye Tracking & 20 & Video clips & 6 basic emotions, Neutral \\
DREAMER \cite{katsigiannis2018dreamer} & EEG, ECG & 23 & Film clips & Valence, Arousal, Dominance \\
EAV \cite{lee2024eav} & EEG, Audio, Video & 42 & Scripted conversations & 5 emotions \\
AMIGOS \cite{miranda2021amigos} & EEG, ECG, GSR, Video & 40 & Short/long videos & Valence, Arousal, Control \\
AffectNet \cite{mollahosseini2019affectnet} & Facial Images & 60 & Real-world images & 7 expressions, Valence-Arousal \\
K-EmoCon \cite{park2020kemocon} & EEG, ECG, Physiological, Video & 32 & Debates & Continuous emotions \\
WESAD \cite{schmidt2018wesad} & BVP, ECG, EDA, EMG, Resp, Temp, Acc & 15 & Stress tasks & Neutral, Stress, Amusement \\
MAHNOB-HCI \cite{soleymani2012multimodal} & Video, Audio, Eye Gaze, ECG, EMG, EOG, GSR & 27 & Videos, Images & Arousal, Valence, Dominance \\
ASCERTAIN \cite{subramanian2018ascertain} & EEG, ECG, GSR & 58 & Emotional videos & Valence, Arousal, Engagement \\
MGEED \cite{wang2024mgeed} & EEG, ECG, OMG, Video, Depth Maps & 17 & Emotional videos & Valence, Arousal, 6 emotions \\
DEAP \cite{koelstra2012deap} & EEG, GSR, Facial Videos & 32 & Music videos & Valence, Arousal, Liking \\
% EmoSense \cite{smetanin2019emosense} & Textual Conversations & N/A & Dialogue data & Happy, Sad, Angry \\
SEMAINE \cite{mollahosseini2019affectnet} & Facial Videos, Audio & 150 & Dyadic conversations & Valence, Arousal \\
\rowcolor{lightgray}
\textbf{AFFEC} & EEG, Eye Tracking, GSR, Body temperature, Video, Personality &71 & Dialogue \& Facial videos  & Felt, Perceived Emotions \\
\hline
\end{tabular}
\end{table*}

\section{AFFEC Dataset Design and Collection}
\label{sec-dataset}

\subsection{Data Access and Structure}
\label{sec:data_access}

The AFFEC dataset is openly available on Zenodo under at 
\url{https://doi.org/10.5281/zenodo.14794876} \cite{affec2025}. 
It follows the Brain Imaging Data Structure (BIDS) \cite{gorgolewski2016brain} convention and organizes each participant’s recordings (EEG, eye tracking, GSR, facial videos, and behavioural logs) into \texttt{sub-<subject\_id>} folders with corresponding annotation files. At the root level, users will find:

\begin{itemize}
    \item \texttt{dataset\_description.json} --- High-level dataset metadata.
    \item \texttt{participants.tsv} and \texttt{participants.json} --- Demographic data and Big Five personality scores.
    \item \texttt{task-fer\_events.json} --- Global event annotations for the emotion-recognition (FER) task.
    \item \texttt{README.md} --- Detailed documentation on file formats, usage, and data collection.
\end{itemize}

Each subject folder is further divided into sub-directories such as \texttt{eeg/}, \texttt{beh/}, and \texttt{events/}:

\begin{itemize}
    \item \textbf{EEG Data (\texttt{eeg/}):} \texttt{.edf} files sampled at 256\,Hz from 63 channels (10--20 layout), accompanied by JSON sidecar files containing channel information.
    \item \textbf{Eye-Tracking, GSR, and Facial Data (\texttt{beh/}):} JSON or TSV files capturing gaze coordinates, pupil diameter, galvanic skin response, and facial action units.
    \item \textbf{Event Files (\texttt{*\_events.tsv}):} Trial timing, stimulus onsets/offsets, and emotional labels for perceived and felt emotions.
\end{itemize}

All recordings comply with the BIDS specification, ensuring compatibility with common neuroimaging and physiological signal-processing tools. 
Researchers can download the dataset from Zenodo at the link above.

In addition, code for preprocessing, analysis, and baseline modeling is available at 
\url{https://github.com/itubrainlab/AFFEC/tree/main}. 
The GitHub repository includes usage guidelines and sample notebooks, enabling quick integration of AFFEC data into diverse workflows.

\subsection{Participants}
We recruited 73 participants (21 females; mean age $27.4 \pm 6$ years) spanning high-school through PhD educational levels. All participants had normal or corrected-to-normal vision and were neurotypical. Informed consent was obtained from each participant in accordance with institutional ethical standards. 

\subsection{Experimental Design and Procedure}
To simulate the listening component of a conversational setting, participants were exposed to dynamic, emotionally expressive video clips. A total of 88 trials (4 practice trials plus 84 main trials) were presented in random order. We selected 84 video clips from the CREMA-D dataset~\cite{cao2014crema}, covering 91 actors (48 males, 43 females) aged 20--74 who portrayed six basic emotions (Anger, Disgust, Fear, Happy, Neutral, Sad) at varying intensities. This selection balanced emotional content and actor demographics to ensure broad coverage and generalizability.

\begin{figure}[htb]
    \centering
    \includegraphics[width=0.3\textwidth]{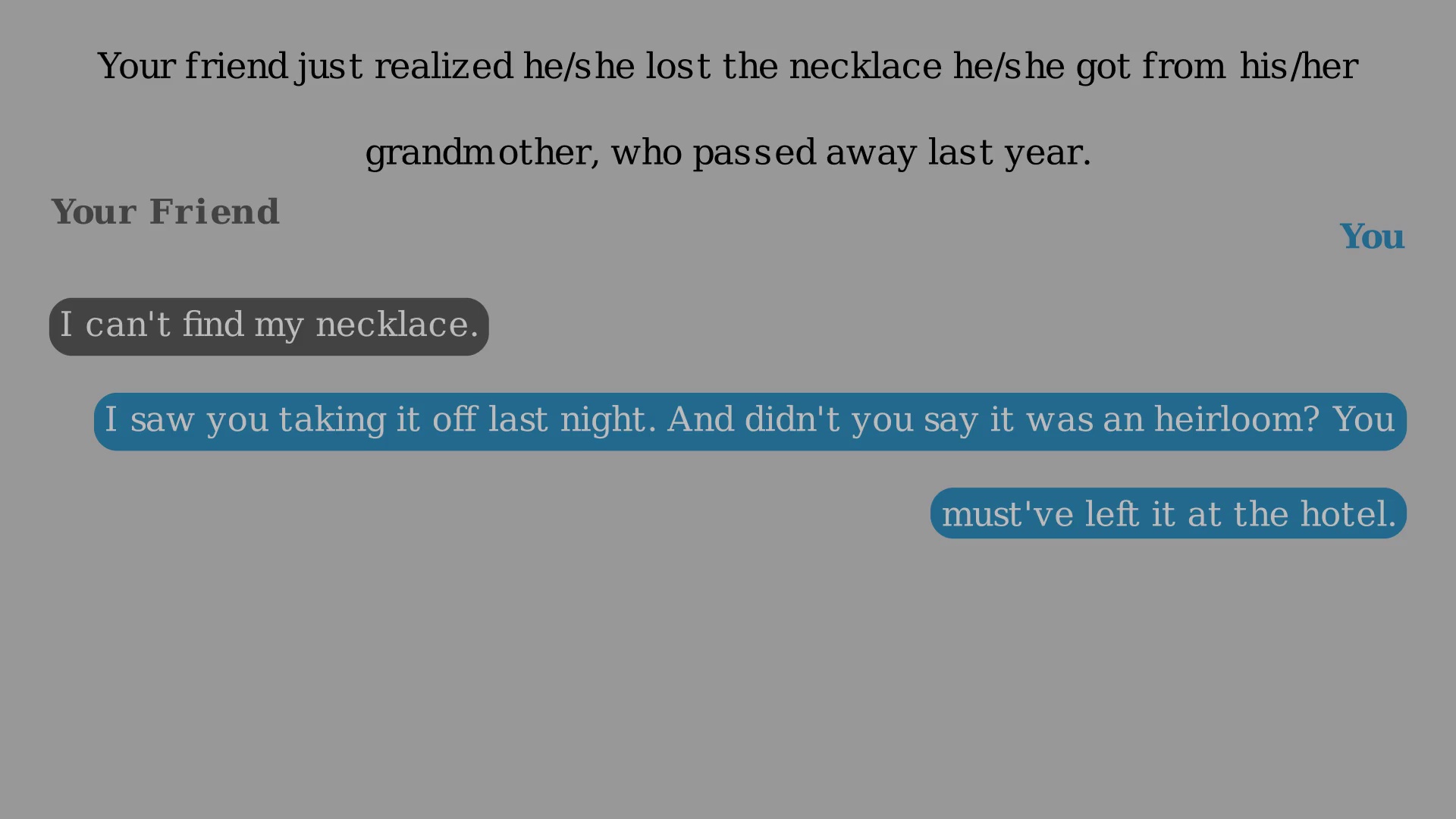}
    \caption{Example scenario presented before the video stimulus. Such contextual prompts prime emotional responses akin to natural interactions.}
    \label{fig:scenario}
\end{figure}

\begin{figure}[htb]
    \centering
    \includegraphics[width=0.2\textwidth]{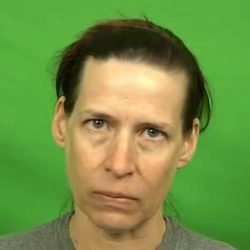}
    \caption{Representative frame from a CREMA-D video stimulus. Dynamic facial expressions support ecological validity in emotion research.}
    \label{fig:example_video_stimuli}
\end{figure}

To enhance ecological validity, each video was preceded by a brief textual scenario offering context for the emotion depicted (Fig.~\ref{fig:scenario}). These scenarios were designed to prime the emotional responses of participants, mimicking realistic conversational cues. Figure~\ref{fig:example_video_stimuli} shows a sample frame from one of the stimuli.

\begin{figure}[htb]
    \centering
    \includegraphics[width=0.4\textwidth]{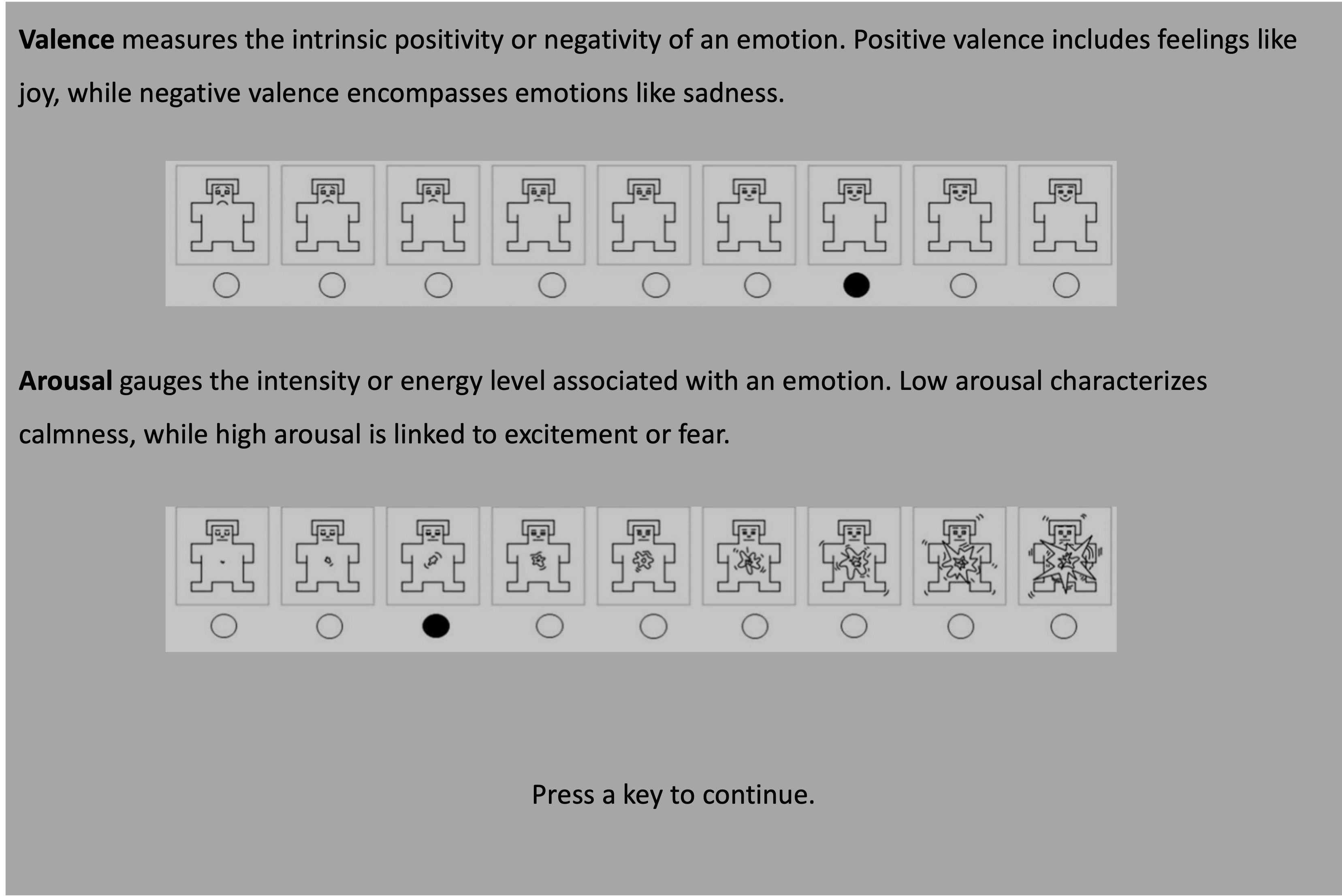}
    \caption{The 9-point scales used for describing the emotional arousal and valence scores.}
    \label{fig:likert_scales}
\end{figure}

After each video, participants provided self-reported ratings for both \emph{perceived} and \emph{felt} emotions. Here, perceived emotions ($E_p$) refer to what the participant believed the video was expressing, whereas felt emotions ($E_f$) capture the participant’s own subjective experience. Ratings were given on 9-point Likert scales for arousal (1 = very calm; 9 = very excited) and valence (1 = very negative; 9 = very positive), as depicted in Fig.~\ref{fig:likert_scales}.

\subsection{Data Collection Modalities}
To enable a comprehensive assessment of emotional responses, multiple data streams were recorded:

\begin{itemize}
    \item \textbf{Eye-Tracking:} Gazepoint GP3 (150 Hz) capturing gaze positions and pupil size, indicative of attention patterns and arousal.
    \item \textbf{EEG:} g.tec hiamp 64-channel system to record neural activity associated with cognitive and emotional processes.
    \item \textbf{GSR:} Shimmer 3 finger-mounted sensors tracking changes in skin conductance reflective of emotional arousal.
    \item \textbf{Facial Movements:} A USB camera recorded participants’ facial expressions throughout each trial.
    \item \textbf{Personality Assessments:} The BFI-44 questionnaire~\cite{john1991big} measured the five personality traits (Openness, Conscientiousness, Extraversion, Agreeableness, and Neuroticism), adding context for individual variability in emotional perception and expression.
\end{itemize}

\begin{figure}[H]
    \centering
    \includegraphics[width=0.45\textwidth]{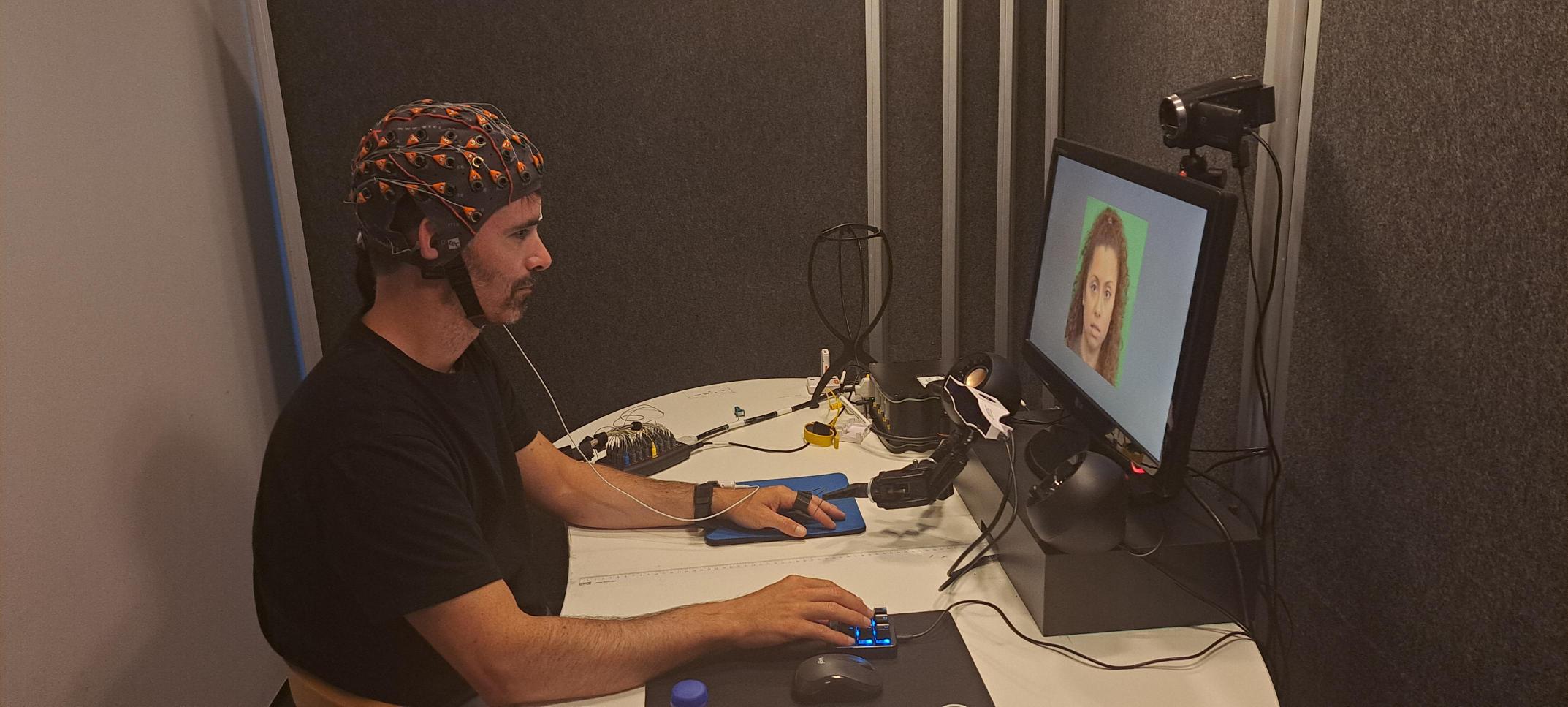}
    \caption{Participant in the experimental setup with EEG, GSR, and eye-tracking sensors, viewing stimuli on a monitor.}
    \label{fig:experiment_setup}
\end{figure}

Figure~\ref{fig:experiment_setup} shows the overall experimental setup, conducted under controlled lighting and temperature to minimize confounds.

\begin{figure}[H]
    \centering
    \includegraphics[width=0.45\textwidth]{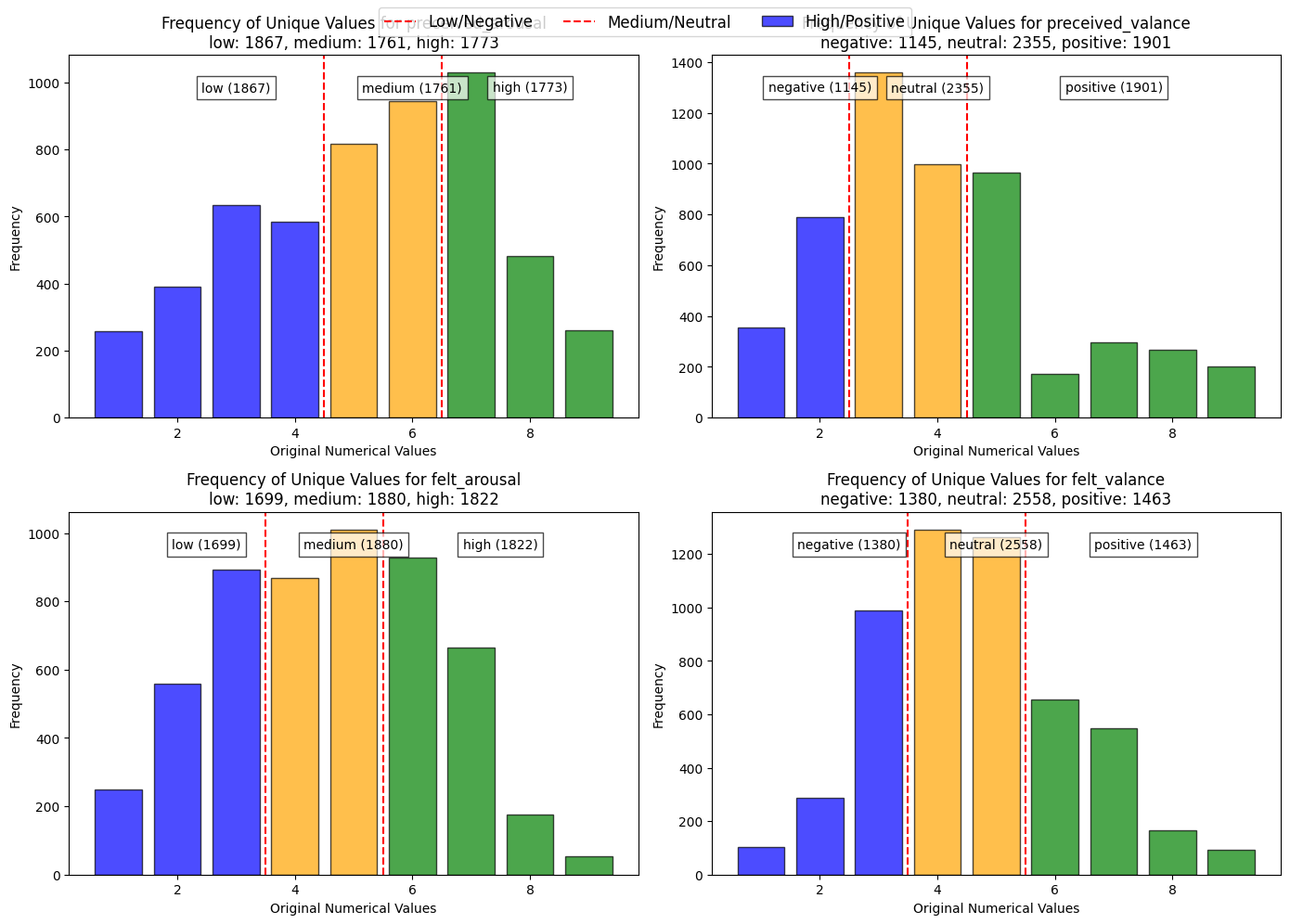}
    \caption{Frequency distributions and corresponding bin boundaries for perceived and felt arousal/valence}
    \label{fig:discretization_visualization}
\end{figure}
\subsection{Pre-Experiment Calibration}
Before the main trials, the following calibrations ensured data integrity:

\begin{itemize}
    \item A 9-point grid for eye-tracker calibration.
    \item EEG and GSR sensor checks for optimal placement and stable signals.
    \item Practice trials to familiarize participants with the ratings and experimental procedure.
\end{itemize}

\subsection{Emotion Labels and Ground Truth}
Participants provided separate ratings for perceived and felt emotions following each video clip. This dual-labeling approach supports nuanced analyses of interactions between the stimulus's expressed emotion ($E_p$) and the participant’s internal affective response ($E_f$). Arousal (1 = very calm; 9 = very excited) and valence (1 = very negative; 9 = very positive) ratings were subsequently discretized into three categories---\emph{low/negative}, \emph{medium/neutral}, and \emph{high/positive}---using an optimal binning strategy. This approach equalizes class distributions (approximately 33\% per category) and facilitates downstream classification. Figure~\ref{fig:discretization_visualization} illustrates the frequency distributions and corresponding bin boundaries for perceived and felt arousal/valence.

\subsubsection{Discretization Process and Insights}
We adopted the following steps:

\begin{itemize}
    \item \textbf{Label Categories:} Continuous ratings mapped into low/negative, medium/neutral, or high/positive.
    \item \textbf{Optimal Binning:} Boundaries were chosen to minimize divergence from a balanced distribution ($\sim33\%$ per class).
    \item \textbf{Visualization:} Figure~\ref{fig:discretization_visualization} shows the frequency distributions with vertical dashed lines marking bin thresholds.
\end{itemize}

Notable findings:
\begin{enumerate}
    \item \textbf{Near-Balanced Classes:} Each of the three categories captured roughly one-third of the data.
    \item \textbf{Positive Valence Bias:} Both perceived and felt valence skewed higher, especially at medium/high levels.
    \item \textbf{Arbitrary Differences:} Slight discrepancies appeared between felt and perceived ratings; for instance, felt valence had a larger neutral category.
    \item \textbf{Model Compatibility:} These categories offer a robust foundation for classification tasks.
\end{enumerate}

\subsubsection{Data Splits and Evaluation}
To ensure consistent comparisons across modalities (e.g., GSR, eye-tracking, EEG), we enforced a uniform data split (60\%/20\%/20\% for train/validation/test) with label stratification. Each experiment used 5-fold cross-validation on the training set for hyperparameter tuning, and the best model was selected based on macro F1-score. We report mean $\pm$ standard deviation across folds to reflect variance.

\subsection{Dataset Statistics}
\label{sec:dataset_stats}
Finally, we summarize the primary dataset features:
\begin{itemize}
    \item \textbf{Eye Tracking:} 16 channels at 62\,Hz (fixations, gaze coordinates, etc.) across 5{,}632 trials.
    \item \textbf{Pupil Data:} 21 channels at 149\,Hz (pupil diameter, eye position), also 5{,}632 trials.
    \item \textbf{Cursor Data:} 4 channels at 62\,Hz (cursor X/Y, state) over 5{,}632 trials.
    \item \textbf{Facial Analysis:} 200+ channels at 40\,Hz (2D/3D landmarks, gaze detection, action units) for 5{,}680 trials.
    \item \textbf{GSR \& Other Physiological:} 40 channels at 50\,Hz (skin conductance, temperature, accelerometer) across 5{,}438 trials.
    \item \textbf{EEG:} 63 channels at 256\,Hz (10--20 scheme; reference: left earlobe) spanning 5{,}632 trials.
    \item \textbf{Self-Annotations:} 5{,}807 trials, each with 4 emotion labels (perceived/felt arousal/valence).
    \item \textbf{Events:} Onset time, duration, trial's type, and emotion labeling markers.
\end{itemize}

This multimodal richness makes AFFEC well suited for advanced affective modeling, offering a more complete view of real-time, face-to-face emotional communication.

\section{Eye Data Analysis}
\label{sec:eye_analysis}

We analyzed eye-tracking data to investigate whether visual attention patterns and pupil dynamics can predict self-reported emotional states. In particular, we focused on four target variables—\textit{perceived arousal}, \textit{perceived valence}, \textit{felt arousal}, and \textit{felt valence}—each discretized into three ordinal categories (low, medium, high) as described in Section~\ref{sec-dataset}.

\subsection{Modeling Approach}
For each target variable, we trained a separate classifier using features derived from the eye-tracking data. The key features extracted per trial include:
\begin{itemize}
    \item \textbf{Gaze Position:} Horizontal and vertical coordinates.
    \item \textbf{Fixation Metrics:} Duration and count of fixations.
    \item \textbf{Pupil Diameter:} Statistical measures capturing pupil size variations.
\end{itemize}

For each trial, we computed summary statistics (mean, standard deviation, minimum, and maximum) over the time-series data for these features. The dataset was then stratified and split into 60\% training, 20\% validation, and 20\% test sets to ensure balanced class distributions. To address class imbalance, class weights were applied during model training.

Random Forest classifiers were selected for their robustness, and model performance was evaluated using 5-fold cross-validation. The evaluation metrics include per-class F1-scores, macro-averaged F1-scores, and overall accuracy.

\subsection{Results}
Table~\ref{tab:multimodal_results} summarises the performance of the best models using eye-tracking features. Overall, the models achieve moderate predictive accuracy for arousal-related measures. However, the prediction of valence—particularly in the high category—remains challenging. These results indicate that while eye-tracking data capture important aspects of visual attention and arousal, additional modalities may be required to robustly predict emotional valence.

\begin{table*}[h!]
\centering
\caption{Classification Performance (F1 Score) Using Eye-Tracking Features (5-Fold Cross-Validation)}
\label{tab:multimodal_results}
\begin{tabular}{lcccc}
\toprule
 & \textbf{Perceived Arousal} & \textbf{Perceived Valence} & \textbf{Felt Arousal} & \textbf{Felt Valence} \\ 
\cmidrule(lr){2-2} \cmidrule(lr){3-3} \cmidrule(lr){4-4} \cmidrule(lr){5-5}
\textbf{Best Model} & LightGBM & RandomForest & ExtraTrees & RandomForest \\ \midrule
\textbf{High} & 0.4059 $\pm$ 0.0324 & 0.0416 $\pm$ 0.0120 & 0.3039 $\pm$ 0.0114 & 0.2696 $\pm$ 0.0565 \\ 
\textbf{Medium} & 0.3070 $\pm$ 0.0368 & 0.4240 $\pm$ 0.0254 & 0.4913 $\pm$ 0.0209 & 0.2752 $\pm$ 0.0409 \\ 
\textbf{Low} & 0.4858 $\pm$ 0.0360 & 0.5697 $\pm$ 0.0101 & 0.6594 $\pm$ 0.0132 & 0.6027 $\pm$ 0.0252 \\[3pt]
\textbf{Macro Avg} & 0.3996 $\pm$ 0.0120 & 0.3451 $\pm$ 0.0116 & 0.4849 $\pm$ 0.0116 & 0.3825 $\pm$ 0.0304 \\ 
\textbf{Accuracy} & 0.4123 $\pm$ 0.0143 & 0.4576 $\pm$ 0.0126 & 0.5501 $\pm$ 0.0135 & 0.4642 $\pm$ 0.0304 \\ 
\bottomrule
\end{tabular}
\end{table*}

% These findings indicate that **gaze and pupil dynamics capture some aspects of emotional arousal but are less effective for valence prediction**. Future work may explore the integration of additional physiological signals, such as GSR and EEG, to improve classification performance.

\section{EEG Analysis}
\label{sec:eeg_analysis}

\subsection{Preprocessing}
To standardise data recorded under varying configurations, minimal preprocessing was applied using \texttt{MNE v. 1.8.0.} All EEG signals were resampled to 200 Hz and bandpass filtered between 1 and 80 Hz. Epochs were extracted from video onset with a fixed duration of 3 seconds. For coarse artefact removal, values above 100 µV and below -100 µV were clipped, and a channel-wise standard scaling was then performed.

\subsection{Modeling Approach}
To demonstrate the suitability of AFFEC's EEG data for emotion recognition for more advanced modelling approaches, we employed a baseline deep learning model using FBSCPNet \cite{schirrmeisterDeepLearningConvolutional2017}, a very popular ConvNet. This model consists of two 1-dimensional convolutional layers, one for the temporal dimension and one for the spatial (channels), with 40 kernels each. For regularisation batch normalization and dropout of 0.5 is used, followed by the output classification layer. All hyperparameters were kept as presented in the original paper and for all four targets. The model was then trained for 5000 epochs and tested on a 5-fold cross-validation split. We stress here that no hyperparameter tuning was performed, and no upsampling was applied, contrary to common practices in EEG emotion recognition to preserve the natural data distribution. Positive results from this hands-off approach show that the data can be used effectively with as much of an off-the-shelf method as can be found in deep learning modeling of EEG signals. Future work may explore data augmentation techniques (e.g., adding noise or windowing) to further boost performance.

\subsection{Results}
Table~\ref{tab:EEGresults} shows the classification performance of our CNN model on the four target variables, evaluated via 5-fold cross-validation. All metrics are well above the chance level (0.33). Notably, the model achieved its best performance for \textit{Felt Arousal} (F1 = 0.50) while \textit{Perceived Valence} scored the lowest (F1 = 0.40). These results show that the EEG data collected can indeed be used to successfully classify the labels in a robust manner with minimal data preprocessing and model tuning.

\begin{table*}[h!]
\centering
\caption{Classification Performance (F1 Score) CNN on EEG signals (5-Fold Cross-Validation)}
\label{tab:EEGresults}
\begin{tabular}{lcccc}
\toprule
& \multicolumn{1}{c}{\textbf{Perceived Arousal}} & \multicolumn{1}{c}{\textbf{Perceived Valence}} & \multicolumn{1}{c}{\textbf{Felt Arousal}} & \textbf{Felt Valence} \\ 
\cmidrule(lr){2-2} \cmidrule(lr){3-3} \cmidrule(lr){4-4} \cmidrule(lr){5-5}
\textbf{High} & 0.4650 ± 0.0249 & 0.3744 ± 0.0243 & 0.5347 ± 0.0166 & 0.4004 ± 0.0269\\
\textbf{Medium} & 0.3840 ± 0.0222 & 0.4765 ± 0.0066 & 0.4490 ± 0.0165 & 0.5503 ± 0.0376\\
\textbf{Low} & 0.4544 ± 0.0229 & 0.3374 ± 0.0141 & 0.5049 ± 0.0222 & 0.3794 ± 0.0467\\
\textbf{Macro Avg} & 0.4345 ± 0.0067 & 0.3961 ± 0.0106 & 0.4962 ± 0.0135 & 0.4434 ± 0.0177\\
\textbf{Accuracy} & 0.4368 ± 0.0068 & 0.4099 ± 0.0106 & 0.4972 ± 0.0122 & 0.4674 ± 0.0176\\
\bottomrule
\end{tabular}
\end{table*}

\section{GSR Signal Analysis}
\label{sec:gsr_analysis}

Galvanic Skin Response (GSR) signals provide valuable physiological indices of emotional arousal. In this work, we leverage GSR data recorded during exposure to emotionally expressive face videos, along with participants’ self-reported perceived and felt emotions. Our aim is to assess how GSR-derived features relate to both categorical emotional stimuli and continuous affect measures.

\subsection{Data Preprocessing and Feature Extraction}
\begin{figure}[H]
    \centering
    \includegraphics[width=0.45\textwidth]{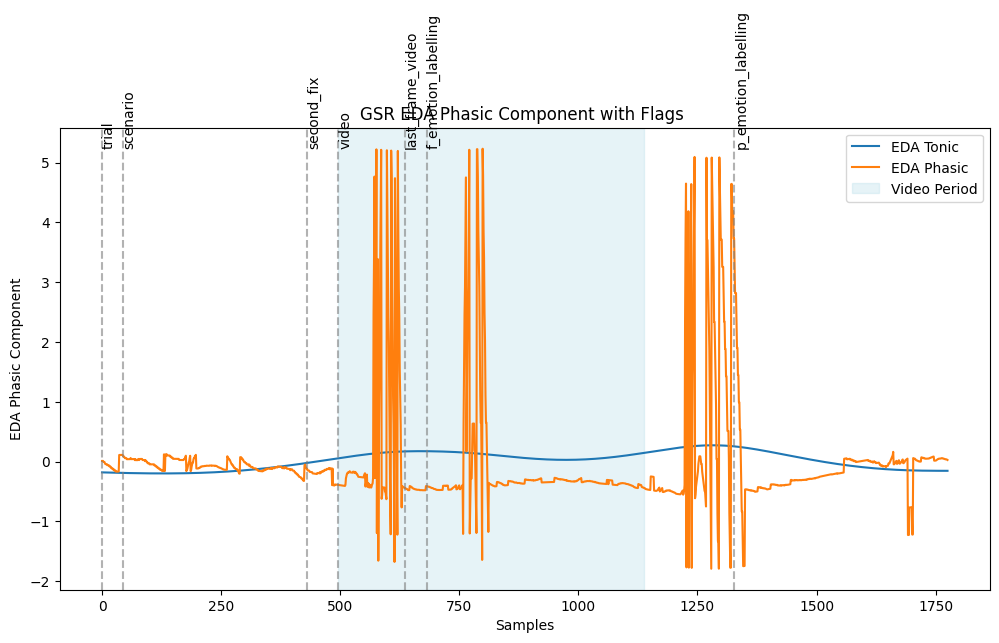}
    \caption{Decomposition of GSR signals into phasic and tonic components.}
    \label{fig:gsr_signal}
\end{figure}

Raw GSR signals were processed using the \texttt{neurokit} module, which decomposes each signal into phasic (rapid fluctuations due to sudomotor activity) and tonic (slowly varying baseline) components. For each trial (from stimulus onset to 10 seconds post-offset), we extracted Skin Conductance Response (SCR) peaks and computed the following metrics:
\begin{itemize}
    \item \textbf{Number of Peaks}
    \item \textbf{SCR Onsets:} mean, median, minimum, maximum, and standard deviation.
    \item \textbf{SCR Amplitude:} mean, median, minimum, maximum, and standard deviation.
    \item \textbf{SCR Height:} mean, median, minimum, maximum, and standard deviation.
    \item \textbf{SCR Rise Time:} mean, median, minimum, maximum, and standard deviation.
    \item \textbf{SCR Recovery Time:} mean, median, minimum, maximum, and standard deviation.
\end{itemize}
These metrics form a comprehensive feature set (see Table~\ref{tab:peak_features}), and Figure~\ref{fig:gsr_signal} illustrates an example of the GSR signal decomposition.

\begin{table}[h!]
\centering
\caption{Extracted SCR Peak Metrics Per Trial}
\label{tab:peak_features}
\begin{tabular}{l}
\toprule
\textbf{Peak Metrics} \\ \midrule
Number of Peaks \\
SCR\_Onsets (mean, median, min, max, STD) \\
SCR\_Amplitude (mean, median, min, max, STD) \\
SCR\_Height (mean, median, min, max, STD) \\
SCR\_RiseTime (mean, median, min, max, STD) \\
SCR\_RecoveryTime (mean, median, min, max, STD) \\ 
\bottomrule
\end{tabular}
\end{table}

\subsection{Classification of Self-Reported Emotional Responses}
Beyond examining statistical relationships, we evaluated the predictive capacity of the GSR features for classifying self-reported emotional responses into three ordinal categories (low, medium, high) for each of the four targets: perceived arousal, perceived valence, felt arousal, and felt valence.

\subsubsection{Experimental Setup}
Continuous ratings for each target were discretised into the three classes as described in Section~\ref{sec-dataset}. After removing missing values and applying this mapping, the extracted SCR metrics were used as input features. We applied an 80/20 train-test split with stratification and trained a \texttt{LGBMClassifier} with class weights to address the imbalance. Model performance was evaluated using precision, recall, and F1-scores, averaged over 5-fold cross-validation.

\subsubsection{Results Using Only GSR Peak Metrics}
Table~\ref{tab:results_gsr_only} presents the classification performance (mean $\pm$ standard deviation) for each target. Although the F1 scores remain modest, the performance for targets such as perceived arousal and felt valence is slightly higher compared to the others.

\begin{table*}[h!]
\centering
\caption{Classification Performance (F1 Score) Using GSR Peak Metrics (5-Fold Cross-Validation)}
\label{tab:results_gsr_only}
\begin{tabular}{lcccc}
\toprule
 & \textbf{Perceived Arousal} & \textbf{Perceived Valence} & \textbf{Felt Arousal} & \textbf{Felt Valence} \\ 
\cmidrule(lr){2-2} \cmidrule(lr){3-3} \cmidrule(lr){4-4} \cmidrule(lr){5-5}
\textbf{Class} & F1 Score (Mean $\pm$ Std) & F1 Score (Mean $\pm$ Std) & F1 Score (Mean $\pm$ Std) & F1 Score (Mean $\pm$ Std) \\ 
\textbf{Best Model} & LightGBM & RandomForest & RandomForest & RandomForest \\ \midrule
High   & 0.3719 $\pm$ 0.0343 & 0.4055 $\pm$ 0.0168 & 0.5720 $\pm$ 0.0104 & 0.5644 $\pm$ 0.0042 \\
Low    & 0.4453 $\pm$ 0.0330 & 0.2227 $\pm$ 0.0339 & 0.3558 $\pm$ 0.0187 & 0.2596 $\pm$ 0.0232 \\
Medium & 0.3056 $\pm$ 0.0398 & 0.4743 $\pm$ 0.0115 & 0.1248 $\pm$ 0.0175 & 0.2186 $\pm$ 0.0174 \\[3pt]
\textbf{Accuracy} & 0.3834 $\pm$ 0.0127 & 0.4016 $\pm$ 0.0109 & 0.4463 $\pm$ 0.0111 & 0.4165 $\pm$ 0.0056 \\
\textbf{Macro Avg} & 0.3743 $\pm$ 0.0177 & 0.3675 $\pm$ 0.0155 & 0.3509 $\pm$ 0.0097 & 0.3475 $\pm$ 0.0088 \\
\bottomrule
\end{tabular}
\end{table*}

\subsubsection{Summary}
GSR-derived features demonstrate sensitivity to emotional arousal, yet the complexity and subjectivity inherent in self-reported affect result in modest classification performance when using GSR data alone. Future work may improve predictive accuracy by integrating additional modalities or employing more advanced modelling techniques.

\section{Facial Videos Analysis}
\label{sec:facial_analysis}

In addition to physiological and eye-tracking data, we examined participants' facial expressions recorded during the experiment. Our analysis focuses on quantifying the alignment between participants' facial expressions and the emotional content of the stimuli, as well as assessing the utility of facial features for emotion recognition.

\subsection{Variables and Measures}
To investigate the relationship between facial expressions and emotional processing, we defined the following measures:
\begin{itemize}
    \item \textbf{Sympathy:} The Euclidean distance between perceived and felt emotion ratings in the valence–arousal space. Lower sympathy values indicate closer alignment between the internal (felt) and external (perceived) emotional states.
    \item \textbf{Facial Mimicry Similarity:} The average Dynamic Time Warping (DTW) score computed across all Action Units (AUs) within each trial. Higher DTW scores reflect stronger temporal alignment between participants' facial expressions and those exhibited in the stimuli.
    \item \textbf{Emotion Recognition Performance:} Evaluated using the Davies–Bouldin Index (DBI) to assess the compactness and separability of emotion clusters derived from facial features. Lower DBI values suggest better-defined clusters.
\end{itemize}

\subsection{Results}
Table~\ref{tab:multimodal_results} presents the classification performance (F1 scores) obtained using facial action unit features for predicting four target emotional variables: perceived arousal, perceived valence, felt arousal, and felt valence. The results, computed via 5-fold cross-validation, indicate that facial features contribute valuable information for emotion recognition, although performance varies across targets. Notably, the best model for predicting felt valence (RandomForest) achieved relatively high F1 scores compared to the other targets.

\begin{table*}[h!]
\centering
\caption{Classification Performance (F1 Score) Using Facial Action Unit Features (5-Fold Cross-Validation)}
\label{tab:multimodal_results}
\begin{tabular}{lcccc}
\toprule
 & \textbf{Perceived Arousal} & \textbf{Perceived Valence} & \textbf{Felt Arousal} & \textbf{Felt Valence} \\ 
\cmidrule(lr){2-2} \cmidrule(lr){3-3} \cmidrule(lr){4-4} \cmidrule(lr){5-5}
\textbf{Best Model} & LightGBMXT & NeuralNetFastAI & NeuralNetFastAI & RandomForest \\ \midrule
\textbf{High}   & 0.4436 $\pm$ 0.0248 & 0.2292 $\pm$ 0.0307 & 0.2344 $\pm$ 0.0362 & 0.4516 $\pm$ 0.0594 \\ 
\textbf{Medium} & 0.2986 $\pm$ 0.0267 & 0.4316 $\pm$ 0.0364 & 0.4778 $\pm$ 0.0348 & 0.2481 $\pm$ 0.0583 \\ 
\textbf{Low}    & 0.4860 $\pm$ 0.0167 & 0.5660 $\pm$ 0.0134 & 0.6176 $\pm$ 0.0196 & 0.6584 $\pm$ 0.0059 \\[3pt]
\textbf{Macro Avg} & 0.4094 $\pm$ 0.0039 & 0.4089 $\pm$ 0.0174 & 0.4433 $\pm$ 0.0213 & 0.4527 $\pm$ 0.0313 \\ 
\textbf{Accuracy}  & 0.4229 $\pm$ 0.0072 & 0.4771 $\pm$ 0.0033 & 0.5157 $\pm$ 0.0211 & 0.5340 $\pm$ 0.0115 \\ 
\bottomrule
\end{tabular}
\end{table*}

These results suggest that facial video data, when analysed via action unit features, provides meaningful cues for emotion recognition. Integrating these features with other modalities may further enhance performance.

\section{Personality Analysis}
\label{sec:personality_analysis}

Personality traits play a critical role in shaping how individuals perceive, experience, and express emotions. In AFFEC, we measured the Big Five traits—Openness, Conscientiousness, Extraversion, Agreeableness, and Neuroticism—using the BFI-44 questionnaire \cite{john1991big}. This section presents both the main findings of how personality correlates with perceived and felt emotions, as well as the detailed scale checks, normality tests, and correlation matrices that were originally placed in the appendix.

\subsection{Measurement Scales}
Table~\ref{tab:personality_meas_scale} presents the range of possible scores for each of the Big Five traits (scaled from 0 to 50 in our questionnaire) and the corresponding ranges for the emotional response ratings (from 1 to 9 on arousal and valence).

\begin{table}[!h]
\centering
\caption{Measurement Scale for Personality and Emotional Responses}
\label{tab:personality_meas_scale}
\begin{tabular}{l r r}
\toprule
\textbf{Variable} & \textbf{Min} & \textbf{Max} \\
\midrule
Openness            & 27.0 & 49.0 \\
Conscientiousness   & 20.0 & 42.0 \\
Extraversion        & 14.0 & 38.0 \\
Agreeableness       & 15.0 & 45.0 \\
Neuroticism         &  9.0 & 35.0 \\
Perceived Arousal   &  1   &  9   \\
Perceived Valence   &  1   &  9   \\
Felt Arousal        &  1   &  9   \\
Felt Valence        &  1   &  9   \\
\bottomrule
\end{tabular}
\end{table}

\subsection{Normality Tests}
While large-sample data often fail formal tests of normality, we nonetheless conducted Shapiro--Wilk tests on each personality trait and emotional response variable to gauge distribution shapes (Table~\ref{tab:personality_shapiro}). Each $p$-value was below 0.05, which is not unusual given our sample size. We thus relied on nonparametric correlations (Pearson or Spearman, depending on subsequent checks) to assess relationships.

\begin{table}[H]
\centering
\caption{Shapiro--Wilk Normality Test Results}
\label{tab:personality_shapiro}
\begin{tabular}{l r r}
\toprule
\textbf{Variable} & \textbf{W Statistic} & \textbf{$p$-value} \\
\midrule
Openness            & 0.9772 & $1.65\times10^{-29}$ \\
Conscientiousness   & 0.9795 & $3.98\times10^{-28}$ \\
Extraversion        & 0.9702 & $4.42\times10^{-33}$ \\
Agreeableness       & 0.9736 & $1.86\times10^{-31}$ \\
Neuroticism         & 0.9778 & $3.57\times10^{-29}$ \\
Perceived Arousal   & 0.9538 & $2.11\times10^{-39}$ \\
Perceived Valence   & 0.9203 & $4.58\times10^{-48}$ \\
Felt Arousal        & 0.9617 & $1.30\times10^{-36}$ \\
Felt Valence        & 0.9583 & $7.02\times10^{-38}$ \\
\bottomrule
\end{tabular}
\end{table}

\subsection{Personality-Emotion Correlations}
Table~\ref{tab:personality_corr_matrix} provides the complete set of pairwise Pearson correlations between each personality trait and the four emotional response variables (perceived arousal, perceived valence, felt arousal, felt valence). Although many correlations are small, a subset stand out as both significant and theoretically consistent with prior literature.

\begin{table}[H]
\centering
\caption{Full Personality--Emotion Pearson Correlations}
\label{tab:personality_corr_matrix}
\begin{tabular}{l l r r}
\toprule
\textbf{Personality} & \textbf{Emotion} & \textbf{$r$} & \textbf{$p$-value} \\
\midrule
Openness           & Perceived Arousal &  0.0866 & $3.80\times10^{-11}$ \\
Openness           & Perceived Valence &  0.0136 & 0.300 \\
Openness           & Felt Arousal      &  0.1382 & $3.68\times10^{-26}$ \\
Openness           & Felt Valence      &  0.0467 & $3.76\times10^{-4}$ \\
Conscientiousness  & Perceived Arousal & -0.0188 & 0.151 \\
Conscientiousness  & Perceived Valence &  0.0272 & 0.038 \\
Conscientiousness  & Felt Arousal      & -0.0106 & 0.418 \\
Conscientiousness  & Felt Valence      &  0.0604 & $4.05\times10^{-6}$ \\
Extraversion       & Perceived Arousal &  0.0143 & 0.277 \\
Extraversion       & Perceived Valence & -0.0399 & 0.0024 \\
Extraversion       & Felt Arousal      &  0.1220 & $1.04\times10^{-20}$ \\
Extraversion       & Felt Valence      & -0.0016 & 0.904 \\
Agreeableness      & Perceived Arousal &  0.0161 & 0.220 \\
Agreeableness      & Perceived Valence & -0.0009 & 0.947 \\
Agreeableness      & Felt Arousal      &  0.0922 & $1.94\times10^{-12}$ \\
Agreeableness      & Felt Valence      &  0.0826 & $2.97\times10^{-10}$ \\
Neuroticism        & Perceived Arousal & -0.0344 & 0.0087 \\
Neuroticism        & Perceived Valence & -0.0174 & 0.185 \\
Neuroticism        & Felt Arousal      & -0.0678 & $2.34\times10^{-7}$ \\
Neuroticism        & Felt Valence      & -0.0883 & $1.61\times10^{-11}$ \\
\bottomrule
\end{tabular}
\end{table}

\subsection{Key Findings}
Although Table~\ref{tab:personality_corr_matrix} enumerates all of the pairwise correlations, several stand out in size and significance:

\begin{itemize}
    \item \textbf{Openness} is positively correlated with \emph{felt arousal} ($r \approx 0.138, p < 10^{-25}$), suggesting that individuals high in openness tend to experience more intense internal emotional states.
    \item \textbf{Extraversion} also shows a positive association with \emph{felt arousal} ($r \approx 0.122, p < 10^{-20}$).
    \item \textbf{Conscientiousness} and \textbf{Agreeableness} show modest positive correlations with \emph{felt valence}, suggesting a slight tilt toward more positive emotional states in higher scorers.
    \item \textbf{Neuroticism} is negatively correlated with both \emph{felt arousal} and \emph{felt valence}, consistent with a propensity for distress and more negative affect.
\end{itemize}

We collect several of these strongest associations in Table~\ref{tab:main_personality}.

\begin{table}[H]
\centering
\caption{Key Personality--Emotion Correlations (Selected from Table~\ref{tab:personality_corr_matrix})}
\label{tab:main_personality}
\begin{tabular}{l l c c}
\toprule
\textbf{Personality} & \textbf{Emotion} & \textbf{$r$} & \textbf{$p$-value} \\
\midrule
Openness          & Felt Arousal  &  0.138 & $< 10^{-25}$ \\
Extraversion      & Felt Arousal  &  0.122 & $< 10^{-20}$ \\
Conscientiousness & Felt Valence  &  0.060 & $< 10^{-5}$  \\
Agreeableness     & Felt Valence  &  0.083 & $< 10^{-9}$  \\
Neuroticism       & Felt Arousal  & -0.068 & $< 10^{-6}$  \\
Neuroticism       & Felt Valence  & -0.088 & $< 10^{-10}$ \\
\bottomrule
\end{tabular}
\end{table}

\subsection{Implications and Future Directions}
These personality--emotion correlations demonstrate that trait-level differences systematically modulate affective responses. In practice, integrating personality measures into affective computing pipelines can enable more adaptive, user-specific emotion recognition systems. For example, highly conscientious and agreeable users may exhibit more positive valence overall, whereas individuals high in neuroticism may require special modeling to handle more frequent negative affect.

Future work might investigate sub-facets of each Big Five trait (e.g., anxiety vs.\ depression within neuroticism) or adopt mixed-effects models that account for repeated measures. In conjunction with the multimodal signals of AFFEC, these personality insights can help researchers build robust, personalized affective computing models that reflect the interplay of stable traits and transient emotional states.

\section{Multimodal Analysis}
\label{sec:multimodal}

In this section, we evaluate the performance of emotion classification for baseline models using multimodal features. Our approach fuses signals from eye-tracking, facial action units (extracted from facial videos), and GSR. We further examine the effect of incorporating personality characteristics into the multimodal representation. Note that these baselines are based on minimal data processing and straightforward feature extraction.

\subsection{Results with Multimodal Features (Eye, Facial Action Units, GSR)}
Table~\ref{tab:multimodal_results} summarises the 5-fold cross-validation performance when combining eye-tracking, facial action unit, and GSR features. The reported F1 scores (along with standard deviations) are provided for each emotion target.
\begin{itemize}
    \item \textbf{Perceived Arousal:} The best model (LightGBM) achieves a macro-average F1 score of 0.4384 $\pm$ 0.0113, with individual class performance ranging from 0.3408 (medium) to 0.5183 (low).
    \item \textbf{Perceived Valence:} The best model (NeuralNetFastAI) reaches a macro-average F1 score of 0.4230 $\pm$ 0.0150.
    \item \textbf{Felt Arousal:} With XGBoost as the best model, the macro-average F1 score is 0.4619 $\pm$ 0.0119.
    \item \textbf{Felt Valence:} Using NeuralNetTorch, the macro-average F1 score is 0.4527 $\pm$ 0.0159.
\end{itemize}

\begin{table*}[H]
\centering
\caption{Classification Performance (F1 Score) Using Multimodal Features (Eye, Facial Action Units, GSR) (5-Fold Cross-Validation)}
\label{tab:multimodal_results}
\begin{tabular}{lcccc}
\toprule
 & \textbf{Perceived Arousal} & \textbf{Perceived Valence} & \textbf{Felt Arousal} & \textbf{Felt Valence} \\ 
\cmidrule(lr){2-2} \cmidrule(lr){3-3} \cmidrule(lr){4-4} \cmidrule(lr){5-5}
\textbf{Best Model} & LightGBM & NeuralNetFastAI & XGBoost & NeuralNetTorch \\ \midrule
High    & 0.4560 $\pm$ 0.0268 & 0.2780 $\pm$ 0.0458 & 0.2241 $\pm$ 0.0310 & 0.4454 $\pm$ 0.0293 \\ 
Medium  & 0.3408 $\pm$ 0.0227 & 0.4277 $\pm$ 0.0227 & 0.4821 $\pm$ 0.0157 & 0.2664 $\pm$ 0.0574 \\ 
Low     & 0.5183 $\pm$ 0.0316 & 0.5632 $\pm$ 0.0205 & 0.6795 $\pm$ 0.0082 & 0.6462 $\pm$ 0.0221 \\[3pt]
\textbf{Macro Avg} & 0.4384 $\pm$ 0.0113 & 0.4230 $\pm$ 0.0150 & 0.4619 $\pm$ 0.0119 & 0.4527 $\pm$ 0.0159 \\ 
\textbf{Accuracy}  & 0.4477 $\pm$ 0.0143 & 0.4757 $\pm$ 0.0156 & 0.5620 $\pm$ 0.0102 & 0.5334 $\pm$ 0.0195 \\ 
\bottomrule
\end{tabular}
\end{table*}

\subsection{Results with Personality Features}
Table~\ref{tab:multimodal_results_personality} presents the classification performance when personality features are integrated with the multimodal data (eye, facial action units, and GSR). Comparing these results with those in Section~\ref{sec:multimodal}, we observe that:
\begin{itemize}
    \item For \textbf{Perceived Arousal} and \textbf{Perceived Valence}, the macroaverage F1 scores remain comparable (0.4377 $\pm$ 0.0080 and 0.4242 $\pm$ 0.0152, respectively).
    \item For \textbf{Felt Arousal} and \textbf{Felt Valence}, incorporating personality results in modest improvements, yielding macro-average F1 scores of 0.4778 $\pm$ 0.0142 and 0.4600 $\pm$ 0.0194, respectively.
\end{itemize}

\begin{table*}[H]
\centering
\caption{Classification Performance (F1 Score) Using Multimodal Features (Eye, Facial Action Units, GSR, Personality) (5-Fold Cross-Validation)}
\label{tab:multimodal_results_personality}
\begin{tabular}{lcccc}
\toprule
 & \textbf{Perceived Arousal} & \textbf{Perceived Valence} & \textbf{Felt Arousal} & \textbf{Felt Valence} \\ 
\cmidrule(lr){2-2} \cmidrule(lr){3-3} \cmidrule(lr){4-4} \cmidrule(lr){5-5}
\textbf{Best Model} & XGBoost & XGBoost & LightGBMXT & NeuralNetFastAI \\ \midrule
High    & 0.4565 $\pm$ 0.0160 & 0.2317 $\pm$ 0.0249 & 0.2692 $\pm$ 0.0401 & 0.4730 $\pm$ 0.0304 \\ 
Medium  & 0.3619 $\pm$ 0.0168 & 0.4306 $\pm$ 0.0168 & 0.4845 $\pm$ 0.0252 & 0.2938 $\pm$ 0.0270 \\ 
Low     & 0.4945 $\pm$ 0.0212 & 0.6104 $\pm$ 0.0148 & 0.6797 $\pm$ 0.0162 & 0.6133 $\pm$ 0.0249 \\[3pt]
\textbf{Macro Avg} & 0.4377 $\pm$ 0.0080 & 0.4242 $\pm$ 0.0152 & 0.4778 $\pm$ 0.0142 & 0.4600 $\pm$ 0.0194 \\ 
\textbf{Accuracy}  & 0.4415 $\pm$ 0.0085 & 0.5057 $\pm$ 0.0145 & 0.5680 $\pm$ 0.0158 & 0.5139 $\pm$ 0.0218 \\ 
\bottomrule
\end{tabular}
\end{table*}

\subsection{Discussion}
Comparing the performance of each modality and the combined (multimodal) configurations highlights several important takeaways. First, \textbf{eye-tracking} yields moderate success in predicting arousal but struggles more with high valence. This suggests that gaze patterns and pupil dynamics capture aspects of emotional intensity but may not fully account for an emotion’s positivity or negativity. In contrast, \textbf{EEG} shows robust performance across targets, especially for felt arousal, despite minimal artefact removal and no hyperparameter tuning; this underscores the rich signal captured by neural measures. \textbf{GSR} data demonstrate moderate predictive capacity, aligning well with arousal-related states but showing limited effectiveness for valence on their own. Facial action units derived from video recordings offer meaningful cues for distinguishing different emotional states, although accuracy again tends to be higher for arousal.

When combining \textbf{eye-tracking, GSR, and facial} modalities, we observe complementary gains across all emotional targets. The macro-average F1 scores in the multimodal setting range from 0.43 to 0.46, confirming that data fusion enhances recognition and suggesting that each sensor taps unique aspects of emotional expression. Furthermore, introducing \textbf{personality traits} yields modest but consistent improvements, particularly for felt emotions, a finding that underscores the role of stable individual differences in affective experiences.

Because these results are derived from baseline methods with simple feature extraction and no extensive optimisation, they serve primarily as evidence that the AFFEC data set contains discriminative signals for emotion recognition. Future work can build upon these findings by pursuing more advanced feature engineering, refined artefact handling (especially for EEG and GSR), and sophisticated fusion strategies (e.g., attention-based deep learning). These refinements could further uncover the nuanced temporal and cross-modal relationships that underlie human emotional experiences. Overall, the baseline models confirm AFFEC’s potential to drive research in multimodal affective computing, while leaving ample room for improved approaches and deeper insights.

\section{Discussion and Future Directions}
\label{sec:discussion}

Across all modalities considered in this study—eye-tracking, EEG, GSR, and facial video—our data-splitting and evaluation protocols remained consistent. Each modality was divided into 60\% training, 20\% validation, and 20\% testing sets with label stratification, and we trained baseline models under a uniform 5-fold cross-validation regime. These choices aimed to ensure fair modality comparisons, mitigate overfitting, and provide a transparent demonstration of the dataset’s quality.

\subsection{Baseline Findings and Modalities}
The primary motivation behind these baseline experiments was to validate that the AFFEC dataset contains \emph{meaningful} signals for emotion recognition, rather than to optimize model performance fully. Indeed, our results show:

\begin{itemize}
    \item \textbf{Eye-Tracking:} Gaze position, fixation metrics, and pupil dynamics reliably capture aspects of arousal but struggle with valence prediction.
    \item \textbf{EEG:} Even with minimal artefact removal and no hyperparameter tuning, a simple CNN achieves performance well above chance on both perceived and felt emotion categories.
    \item \textbf{GSR:} Phasic and tonic skin conductance data exhibit moderate predictive power, particularly for arousal.
    \item \textbf{Facial Expressions:} Facial action units extracted via OpenFace provide valuable cues; however, accuracy varies across different emotion types.
    \item \textbf{Personality Traits:} Incorporating the Big Five scales yields modest but consistent improvements in classifying felt arousal and valence, highlighting the role of stable individual differences.
\end{itemize}

When multiple data streams (e.g., eye-tracking, GSR, and facial AUs) are fused, performance improves, underscoring the complementary nature of these signals. Yet, \emph{valence} predictions remain more challenging than \emph{arousal}, matching broader trends in affective computing. Overall, these results confirm that AFFEC provides discriminative features across diverse modalities.

\subsection{Building on AFFEC: Advanced Methods and Insights}
Though our baselines use straightforward feature extraction and modeling, two recent efforts have adopted more sophisticated approaches on this dataset:

\begin{itemize}
    \item \textbf{Advanced Facial Mimicry Analysis \cite{j2025exploring}:} One study devoted effort to facial action unit processing employing dynamic time warping (DTW) to align participant AUs with those expressed in the stimuli. By meticulously segmenting and synchronizing the AU patterns, the authors probed how closely the facial expressions of the participants mirrored the video content, revealing richer insights into the interaction between perceived and felt responses. Their results indicated that fine-grained temporal alignment could discriminate subtle mimicry differences, particularly among emotions like fear or anger, and suggested that personality traits modulate the degree of alignment.
    \item \textbf{Personality and Eye-Tracking Fusion\cite{seikavandi2025modelling}:} Another work concentrated on neural-network-based pipelines that integrated personality assessments, high-resolution eye-tracking metrics, and temporal modelling. After extensive data preprocessing—such as more refined pupil-baseline corrections and gaze-region segmentations—the authors used multi-branch architectures to fuse these features. Notably, they reported higher macro-F1 scores than our baselines for both \emph{felt} and \emph{perceived} valence, indicating that deeper personalization and advanced sequence modelling can further enhance emotion prediction.
\end{itemize}

Both endeavors confirm that AFFEC can accommodate more detailed data preprocessing steps (e.g., improved noise filtering, advanced feature engineering) and model designs (e.g., neural networks with multi-branch temporal fusion). They also illustrate how personalization—either via personality traits or by modelling individual temporal patterns—can unlock deeper insights into the differences between internal and externally perceived affect.

\subsection{Key Takeaways and Future Directions}
\begin{itemize}
    \item \textbf{AFFEC as a Foundation:} The dataset’s comprehensive signals across EEG, eye-tracking, GSR, facial analysis, and personality questionnaires make it a valuable platform for exploring more advanced or specialised approaches. 
    \item \textbf{Multimodal Fusion Potential:} Our results and recent extensions highlight the gains from combining modalities. Attention-based architectures, transformer models, or advanced temporal alignment techniques (e.g., DTW) may capture the sequential nature of emotional expression more effectively.
    \item \textbf{Personality-Informed Models:} Empirical gains from personality integration suggest that personalisation is a promising avenue. Future work could refine trait-based or mixed-effects models to handle inter-participant variability more precisely.
    \item \textbf{Extended Scenarios:} AFFEC’s structure also invites expansions such as live, real-time systems or investigations into how contextual factors (like conversation flow) alter the mapping between physiological cues and subjective emotion labels.
\end{itemize}

Ultimately, these baseline evaluations confirm the rich signal content of AFFEC while leaving ample scope for innovative modelling and analysis. By incorporating deeper temporal dynamics, refined personalization, or sophisticated fusion methods, future research can push the boundaries of emotion-aware technology and better characterize the subtleties of face-to-face affective communication.

\section{Conclusion}
AFFEC provides a multimodal dataset uniquely suited for studying face-to-face emotive communication in both perceived and felt contexts. Our baseline analyses, covering EEG, eye-tracking, GSR, facial expression data, and personality measures, confirm the presence of discriminative signals for emotion recognition and underscore that:

\begin{itemize}
    \item \textbf{Arousal is more tractable than valence} across most modalities, though multimodal fusion narrows the gap.
    \item \textbf{Eye-tracking, EEG, and GSR} each offer complementary information about emotional states, and combining them yields notable performance gains.
    \item \textbf{Personality traits} contribute valuable insight into individual differences, especially for felt emotions.
\end{itemize}

The dataset’s comprehensive structure, BIDS formatting, and public availability encourage broad adoption and reproducibility. Moreover, two subsequent studies leveraging AFFEC have demonstrated its flexibility: from employing temporal alignment for in-depth facial mimicry analysis to integrating personality-driven gaze models for improved internal state prediction. We anticipate future work will explore advanced multimodal fusion strategies (e.g., attention mechanisms, graph-based architectures) and examine interactive paradigms or larger demographic samples to further elucidate how personality and context shape emotional communication. Overall, the AFFEC dataset represents a significant step forward for research in affective computing, human-agent interaction, and social robotics, supporting the design of more accurate, adaptive, and empathetic emotion-aware technologies.

\section*{Acknowledgments}
We would like to thank all participants and the research team involved in data collection and processing. This work was supported in part by Pioneer Centre for Artificial Intelligence\footnote{https://www.aicentre.dk/}.

\bibliographystyle{IEEEtran}
\bibliography{references}

%%%%%%%%%%%%%%%%%%%%%%%%%%%%%%%%%%%%%%%%%%%%%%%%%%%%%%%%%%%%%%%%%%%%%%%%%%%%%%%
% Appendix: Detailed Personality Analysis Results
%%%%%%%%%%%%%%%%%%%%%%%%%%%%%%%%%%%%%%%%%%%%%%%%%%%%%%%%%%%%%%%%%%%%%%%%%%%%%%%

\end{document}